\documentclass[aps,prd,twocolumn,superscriptaddress,nofootinbib]{revtex4-1}

\usepackage{latexsym}
\usepackage{amsmath}
\usepackage{amssymb}
\usepackage{amsfonts}
\usepackage{bm}
\usepackage{physics}

\usepackage{color}
\definecolor{purple}{rgb}{0.5,0,0.5}
\definecolor{blue}{rgb}{0.0,0,0.9}
\definecolor{prdblue}{rgb}{0.133,0.118,0.498}
\usepackage[colorlinks=true, pdfstartview=FitV, linkcolor=prdblue, citecolor= prdblue, urlcolor=prdblue]{hyperref}

\usepackage{supertabular} 
\usepackage{placeins}
\usepackage{epsfig}
\usepackage{graphicx}
\usepackage{cancel}

\usepackage{soul} 
\usepackage{color}

\begin{document}

\title{Constituent-quark-model based coupled-channels calculation \\[0.25ex] of the $\mathbf{bb\bar c\bar c}$ and $\mathbf{bc\bar b\bar c}$ tetraquark systems}

\author{P. G. Ortega}
\email[]{pgortega@usal.es}
\affiliation{Departamento de F\'isica Fundamental, Universidad de Salamanca, E-37008 Salamanca, Spain}
\affiliation{Instituto Universitario de F\'isica 
Fundamental y Matem\'aticas (IUFFyM), Universidad de Salamanca, E-37008 Salamanca, Spain}

\author{D. R. Entem}
\email[]{entem@usal.es}
\affiliation{Departamento de F\'isica Fundamental, Universidad de Salamanca, E-37008 Salamanca, Spain}
\affiliation{Instituto Universitario de F\'isica
Fundamental y Matem\'aticas (IUFFyM), Universidad de Salamanca, E-37008 Salamanca, Spain}

\author{F. Fern\'andez}
\email[]{fdz@usal.es}
\affiliation{Instituto Universitario de F\'isica 
Fundamental y Matem\'aticas (IUFFyM), Universidad de Salamanca, E-37008 Salamanca, Spain}

\author{J. Segovia}
\email[]{jsegovia@upo.es}
\affiliation{Departamento de Sistemas F\'isicos, Qu\'imicos y Naturales, Universidad Pablo de Olavide, E-41013 Sevilla, Spain}

\date{\today}

\begin{abstract}
We perform a coupled-channels study of the $bb\bar c\bar c$ and $bc\bar b\bar c$ tetraquark systems in a molecular approach using a constituent quark model which has been widely used to satisfactorily describe a broad range of properties of heavy quark hadron systems, either conventional or exotic.
Within a molecular framework, the interaction in the heavy quark sector is governed by gluon exchange or confinement forces that are inherently color-dependent. While the $B_c B_c$ system contains two identical quarks, enabling stronger interactions via exchange diagrams, the forces in the $B_c \bar{B}_c$ and $(c\bar{c})-(b\bar{b})$ systems are expected to be significantly weaker. Consequently, the theoretical and experimental analysis of $B_c^{(*)} B_c^{(*)}$, $B_c^{(*)} \bar{B}_c^{(*)}$, and charmonium-bottomonium bound structures could play a crucial role in clarifying the dominant mechanisms responsible for the formation of fully-heavy tetraquarks.
For the $bb\bar c\bar c$ tetraquark sector, we find several resonance states with different spin-parity quantum numbers. These resonances are characterized by their proximity, but not too close, to the $B_c^{(\ast)}B_c^{(\ast)}$ thresholds and their large total decay widths, indicating strong decay channels.
In contrast, our analysis of the $bc\bar b\bar c$ tetraquark sector reveals no bound states, virtual states, or resonances; suggesting that tetraquark states of the $(c\bar c)-(b\bar b)$ or $B_c^{(\ast)}\bar B_c^{(\ast)}$ molecular type are unlikely to be formed, within our model assumptions.
\end{abstract}


\maketitle


\section{INTRODUCTION}
\label{sec:intro}

Exploring the spectroscopy, structure, and dynamics of exotic hadrons remains one of the most challenging frontiers in modern physics. In recent years, high-energy experiments have uncovered a rich spectrum of multiquark states that challenge traditional descriptions based on baryonic ($qqq$) or mesonic ($q\bar q$) configurations. The ground breaking discovery of $X(3872)$ by the Belle Collaboration~\cite{Belle:2003nnu} marked a pivotal event, paving the way for the identification of numerous tetraquark states. Among these are the $Z_c(3900)$, $Z_c(4020)$~\cite{BESIII:2013ris, BESIII:2013ouc}, $Z_{cs}(3985)^-$ and $Z_{cs}(4220)^+$~\cite{BESIII:2020qkh, LHCb:2021uow}, which exhibit charmonium-like properties; the $Z_b(10610)$, $Z_b(10650)$ bottomonium-like candidates~\cite{Belle:2011aa, Belle:2015upu}; or those which are clearly exotic states such as the $T_{cc}(3875)^+$~\cite{LHCb:2021vvq, LHCb:2021auc} or the $T_{cs0}(2900)^0$ and $T_{cs1}(2900)^0$~\cite{LHCb:2020pxc} particles. The study of the properties of these exotic hadrons promises to deepen our understanding of the fundamental interactions that govern the subatomic world, transcending the conventional quark compositions.

The fully-heavy tetraquark states $QQ\bar{Q}\bar{Q}$ ($Q=c,\,b$) have recently attracted much attention. In 2017, the CMS Collaboration reported a benchmark measurement of $\Upsilon(1S)$-pair production in $pp$ collisions at $\sqrt{s}$=8 TeV~\cite{CMS:2016liw}. A preliminary analysis of the CMS data shows an excess at 18.4 GeV in the $\Upsilon(1S) \ell^+ \ell^-$ decay channels~\cite{CMS:2016liw, Yi:2018fxo}. This excess, if confirmed by future experiments, may indicate a fully-bottom tetraquark state ($bb\bar{b}\bar{b}$). Besides, a significant peak at $18.2\,\text{GeV}$ was observed in Cu+Au collisions at RHIC~\cite{ANDY:2019bfn} but the LHCb and CMS collaborations~\cite{LHCb:2018uwm, CMS:2020qwa} were not able to confirm it from the $\Upsilon(1S)\mu^+ \mu^-$ invariant mass spectrum. Recently, in the di-$J/\psi$ invariant mass spectrum, a narrow peak at 6.9 GeV, a broad one between 6.2 and 6.8 GeV, and a hint for a possible structure around 7.2 GeV were reported by the LHCb collaboration, which could indicate the existence of fully-charm tetraquarks~\cite{LHCb:2020bwg}. It is then expected that more investigations on the existence of fully-heavy tetraquark states will be performed in the future in, for instance, the LHCb experiment at CERN.

On the theoretical side, the investigation of tetraquark states composed of heavy quarks, either charm ($c$) or bottom ($b$), has been ongoing since the early days of hadron physics, with particular focus on all-charm tetraquark configurations~\cite{Iwasaki:1976cn, Chao:1980dv, Ader:1981db, Badalian:1985es}. Following the discovery of the di-$J/\psi$ structure, theoretical interest in fully heavy four-quark systems has surged, leading to a proliferation of studies exploring various interpretations. These include diquark-antidiquark configurations~\cite{Debastiani:2017msn, Chen:2016jxd, Wang:2019rdo, Bedolla:2019zwg, Giron:2020wpx, Jin:2020jfc, Deng:2020iqw, Faustov:2021hjs, Wang:2020ols, Mutuk:2021hmi}, compact tetraquark models~\cite{Lu:2020cns, Karliner:2020dta, Weng:2020jao, Sonnenschein:2020nwn, Gordillo:2020sgc, Yang:2021hrb}, as well as meson-meson molecular states and coupled-channel effects~\cite{Dong:2020nwy, Albuquerque:2020hio, Guo:2020pvt, Agaev:2023ruu, Niu:2022jqp}.

Despite the substantial number of studies conducted to date, the nature of the $QQ\bar{Q}\bar{Q}$ states remains a subject of debate. While many works predict the existence of bound $cc\bar{c}\bar{c}$ and $bb\bar{b}\bar{b}$ states, others argue that these configurations lie above the corresponding thresholds~\cite{Hughes:2017xie, Chen:2019dvd, Wu:2016vtq, Wang:2019rdo}. Similar conclusions have been drawn for the $bb\bar{c}\bar{c}$ and $bc\bar{b}\bar{c}$ systems, although these mixed-flavor configurations have received comparatively less attention and are often addressed as secondary analyses within studies focused on fully charmed or fully bottomed tetraquarks~\cite{Gordillo:2020sgc, Bedolla:2019zwg, Wang:2019rdo, Faustov:2021hjs, Wu:2016vtq}. For instance, Ref.~\cite{Berezhnoy:2011xn} investigates the $bc\bar{b}\bar{c}$ system using a non-relativistic Schr\"odinger equation within a diquark-antidiquark framework, identifying several candidate states with $J^P = 0^+$, $1^+$, and $2^+$. Similarly, Ref.~\cite{Weng:2020jao} employs an extended chromomagnetic model to explore the compact tetraquark spectrum of $bb\bar{c}\bar{c}$ and $bc\bar{b}\bar{c}$ systems, predicting seven $bc\bar{b}\bar{c}$ candidates, one below the $B_c\bar{B}_c$ threshold and the rest above the $B_c^*\bar{B}_c^*$ threshold, as well as four $bb\bar{c}\bar{c}$ states above the $B_c B_c$ threshold in the positive parity sectors.

In our opinion, more attention should be directed toward the $bb\bar{c}\bar{c}$ and $bc\bar{b}\bar{c}$ sectors, as their properties may offer valuable insights into the underlying nature of fully heavy tetraquark states; namely, whether they are predominantly compact tetraquark configurations or the result of coupled-channel dynamics. Conventional studies of compact tetraquarks typically predict a similar spectrum across these mixed-flavor sectors and the fully charmed or bottom counterparts. In contrast, within a molecular framework, the interaction between two color-singlet mesons in the heavy quark sector is largely governed by gluon exchange or confinement forces that are inherently color-dependent. While the $B_c B_c$ system contains two identical quarks, enabling stronger interactions via exchange diagrams, the forces in the $B_c \bar{B}_c$ and $(c\bar{c})-(b\bar{b})$ systems are expected to be significantly weaker. Consequently, the experimental search for $B_c^{(*)} B_c^{(*)}$, $B_c^{(*)} \bar{B}_c^{(*)}$, and charmonium-bottomonium bound structures could play a crucial role in clarifying the dominant mechanisms responsible for the formation of fully-heavy tetraquarks.

In this work, we explore the possibility of having bound, virtual and resonance molecular states of $bb\bar c\bar c$ and $bc\bar b\bar c$ tetraquark systems, with spin-parity $J^{P}=0^{\pm}$, $1^{\pm}$ and $2^{\pm}$.\footnote{Note here that the analysis for the $cc\bar c \bar c$ and $QQ'\bar q\bar q'$ tetraquark systems were recently published in Refs.~\cite{Ortega:2023pmr,Ortega:2024epk}.} For that purpose we perform a coupled-channels calculation in the framework of the constituent quark model (CQM) proposed in Ref.~\cite{Vijande:2004he}. This model has been extensively used to describe the phenomenology of heavy hadron sectors~\cite{Segovia:2013wma}. In the heavy quark sector, the quark--(anti-)quark potential is given by a screened linear confinement potential and a one-gluon exchange interaction. To find the quark-antiquark bound states we solve the Schr\"odinger equation following Ref.~\cite{Hiyama:2003cu}. To describe the interaction at the meson level, we use the Resonating Group Method (RGM)~\cite{Wheeler:1937zza, Tang:1978zz}, where mesons are considered as quark-antiquark clusters and an effective cluster-cluster interaction emerges from the underlying quark(antiquark) dynamics~\cite{Ortega:2012rs}.

This manuscript is organized as follows. Section~\ref{sec:theory} provides a brief overview of the theoretical framework. Section~\ref{sec:results} primarily focuses on the analysis and discussion of our theoretical findings. Lastly, in Sec.~\ref{sec:summary}, we present a summary of our work and draw some conclusions based on the obtained results.

\section{THEORETICAL FORMALISM}
\label{sec:theory} 

\subsection{Naive quark model}

The main components of our constituent quark model (CQM) in the fully heavy quark sectors are the perturbative one-gluon exchange (OGE) interaction and a non-perturbative confining term~\cite{Vijande:2004he}.

One-gluon fluctuations around the instanton vacuum are taken into account through the $qqg$ coupling~\cite{Diakonov:1985eg, Diakonov:2002fq}
\begin{eqnarray}
{\mathcal L}_{qqg} &=& i\sqrt{4\pi\alpha_s} \bar \psi \gamma_\mu G^\mu_c
\lambda^c \psi,
\label{Lqqg}
\end{eqnarray}
with $\lambda^c$ being the $SU(3)$ color matrices and $G^\mu_c$ the gluon field. The different terms of the potential derived from the Lagrangian shown above contain central, tensor, and spin-orbit contributions; they are given by~\cite{Segovia:2013wma}
\begin{widetext}
\begin{align}
V_{\text{OGE}}^{\text{C}}(\vec{r}_{ij})=
& \frac{1}{4}\alpha_{s}(\vec{\lambda}_{i}^{c}\cdot
\vec{\lambda}_{j}^{c})\left[ \frac{1}{r_{ij}}-\frac{1}{6m_{i}m_{j}} 
(\vec{\sigma}_{i}\cdot\vec{\sigma}_{j}) 
\frac{e^{-r_{ij}/r_{0}(\mu_{ij})}}{r_{ij}r_{0}^{2}(\mu_{ij})}\right], \nonumber \\
V_{\rm OGE}^{\rm T}(\vec{r}_{ij})= &-\frac{1}{16}\frac{\alpha_{s}}{m_{i}m_{j}}
(\vec{\lambda}_{i}^{c}\cdot\vec{\lambda}_{j}^{c})\left[ 
\frac{1}{r_{ij}^{3}}-\frac{e^{-r_{ij}/r_{g}(\mu_{ij})}}{r_{ij}}\left( 
\frac{1}{r_{ij}^{2}}+\frac{1}{3r_{g}^{2}(\mu_{ij})}+\frac{1}{r_{ij}r_{g}(\mu_{ij})}\right) \right] S_{ij}, \nonumber \\
 V_{\rm OGE}^{\rm SO}(\vec{r}_{ij}) = &
-\frac{1}{16}\frac{\alpha_{s}}{m_{i}^{2}m_{j}^{2}}(\vec{\lambda}_{i}^{c}\cdot
\vec{\lambda}_{j}^{c})\left[\frac{1}{r_{ij}^{3}}-\frac{e^{-r_{ij}/r_{g}(\mu_{ij})}}{r_{ij}^{3}} \left(1+\frac{r_{ij}}{r_{g}(\mu_{ij})}\right)\right] \times \nonumber \\
&
\times \left[((m_{i}+m_{j})^{2}+2m_{i}m_{j})(\vec{S}_{+}\cdot\vec{L})+(m_{j}^{2}
-m_{i}^{2}) (\vec{S}_{-}\cdot\vec{L}) \right],
\end{align}
\end{widetext}
where $\vec{S}_{\pm}=\frac{1}{2}(\vec{\sigma}_{i}\,\pm\,\vec{\sigma}_{j})$. Besides, 
$r_{0}(\mu_{ij})=\hat{r}_{0}\frac{\mu_{nn}}{\mu_{ij}}$ and
$r_{g}(\mu_{ij})=\hat{r}_{g}\frac{\mu_{nn}}{\mu_{ij}}$ are regulators which depend on
$\mu_{ij}$, the reduced mass of the $q\bar{q}$ pair. The contact term of the
central potential has been regularized as
\begin{equation}
\delta(\vec{r}_{ij})\sim\frac{1}{4\pi r_{0}^{2}}\frac{e^{-r_{ij}/r_{0}}}{r_{ij}} \,.
\end{equation}

The wide energy range needed to provide a consistent description of light, strange and heavy mesons requires an effective scale-dependent strong coupling constant. We use the frozen coupling constant of Ref.~\cite{Vijande:2004he, Segovia:2008zza}
\begin{equation}
\alpha_{s}(\mu_{ij})=\frac{\alpha_{0}}{\ln\left( 
\frac{\mu_{ij}^{2}+\mu_{0}^{2}}{\Lambda_{0}^{2}} \right)},
\end{equation}
in which $\mu_{ij}$ is the reduced mass of the $q\bar{q}$ pair and $\alpha_{0}$, $\mu_{0}$ and $\Lambda_{0}$ are parameters of the model determined by a global fit to the meson spectra.

Confinement is one of the crucial aspects of QCD. Color charges are confined inside hadrons. It is well known that multi-gluon exchanges produce an attractive linearly rising potential proportional to the distance between infinite-heavy quarks~\cite{Bali:2000gf}. However, sea quarks are also important ingredients of the strong interaction dynamics that contribute to the screening of the rising potential at low momenta and eventually to the breaking of the quark-antiquark binding string~\cite{Bali:2005fu}. Our model tries to mimic this behaviour using the following expression~\cite{Segovia:2013wma}:
\begin{widetext}
\begin{align}
V_{\rm CON}^{\rm C}(\vec{r}_{ij}) =& \left[ -a_{c}(1-e^{-\mu_{c}r_{ij}})+\Delta
\right] (\vec{\lambda}_{i}^{c}\cdot\vec{\lambda}_{j}^{c}), \nonumber\\
V_{\rm CON}^{\rm SO}(\vec{r}_{ij})= &
-\left(\vec{\lambda}_{i}^{c}\cdot\vec{\lambda}_{j}^{c} \right) 
\frac{a_{c}\mu_{c}e^{-\mu_{c}r_{ij}}}{4m_{i}^{2}m_{j}^{2}r_{ij}}\left[((m_{i}^{2
}+m_{j}^{2})(1-2a_{s}) +4m_{i}m_{j}(1-a_{s}))(\vec{S}_{+}\cdot\vec{L})\right.
\nonumber \\
&
\left. +(m_{j}^{2}-m_{i}^{2})(1-2a_{s})(\vec{S}_{-}\cdot\vec{L}) \right], 
\end{align}
\end{widetext}
where $a_{c}$ and $\mu_{c}$ are model parameters. At short distances this potential presents a linear behavior with an effective confinement strength, $\sigma=-a_{c}\,\mu_{c}\,(\vec{\lambda}^{c}_{i}\cdot \vec{\lambda}^{c}_{j})$, while it becomes constant at large distances. This type of potential shows a threshold defined by
\begin{equation}
V_{\rm thr} = \{-a_{c}+\Delta\}(\vec{\lambda}^{c}_{i}\cdot \vec{\lambda}^{c}_{j}) \,.
\end{equation}
No quark-antiquark bound states can be found for energies higher than this threshold. The system suffers a transition from a colour string configuration between two static colour sources into a pair of static mesons due to the breaking of the colour flux-tube and the most favoured subsequent decay into hadrons.

Among the different methods to solve the Schr\"odinger equation in order to  find the quark-antiquark bound states, we use the Gaussian Expansion Method~\cite{Hiyama:2003cu} because it provides sufficient accuracy and simplifies the subsequent evaluation of the required matrix elements. This procedure yields the radial wave function solution of the Schr\"odinger equation as an expansion in terms of basis functions
\begin{equation}
R_{\alpha}(r)=\sum_{n=1}^{n_{max}} c_{n}^\alpha \phi^G_{nl}(r),
\end{equation} 
where $\alpha$ refers to the channel quantum numbers. The coefficients, $c_{n}^\alpha$, and the eigenvalue, $E$, are determined from the Rayleigh-Ritz variational principle
\begin{equation}
\sum_{n=1}^{n_{max}} \left[\left(T_{n'n}^\alpha-EN_{n'n}^\alpha\right)
c_{n}^\alpha+\sum_{\alpha'}
\ V_{n'n}^{\alpha\alpha'}c_{n}^{\alpha'}=0\right],
\end{equation}
where $T_{n'n}^\alpha$, $N_{n'n}^\alpha$ and $V_{n'n}^{\alpha\alpha'}$ are the  matrix elements of the kinetic energy, the normalization and the potential,  respectively. $T_{n'n}^\alpha$ and $N_{n'n}^\alpha$ are diagonal whereas the mixing between different channels is given by $V_{n'n}^{\alpha\alpha'}$.

Following Ref.~\cite{Hiyama:2003cu}, we employ Gaussian trial functions with ranges in geometric progression. This enables the optimization of ranges employing a small number of free parameters. Moreover, the geometric progression is dense at short distances, so that it allows the description of the dynamics mediated by short range potentials. The fast damping of the Gaussian tail is not a problem, since we can choose the maximal range much longer than the hadronic size.

\begin{table}[!t]
\caption{\label{tab:parameters} Quark model parameters.}
\begin{ruledtabular}
\begin{tabular}{lrr}
Quark masses & $m_{c}$ (MeV) & $1763$ \\
 		     & $m_{b}$ (MeV) & $5110$ \\[2ex]
OGE & $\alpha_{0}$ & $2.118$ \\
     & $\Lambda_{0}$ $(\mbox{fm}^{-1})$ & $0.113$ \\
     & $\mu_{0}$ (MeV) & $36.976$ \\
     & $\hat{r}_{0}$ (fm) & $0.181$ \\
     & $\hat{r}_{g}$ (fm) & $0.259$ \\[2ex]
Confinement & $a_{c}$ (MeV) & $507.4$ \\
	       & $\mu_{c}$ $(\mbox{fm}^{-1})$ & $0.576$ \\
	       & $\Delta$ (MeV) & $184.432$ \\
	       & $a_{s}$ & $0.81$ \\
\end{tabular}
\end{ruledtabular}
\end{table}

Table~\ref{tab:parameters} shows the involved model parameters fitted over all meson spectra~\cite{Vijande:2004he}, updated in Refs.~\cite{Segovia:2008zza, Segovia:2008zz}.
The model has been successfully applied to the description of the spectra of charmonium~\cite{Segovia:2008zz}, bottomonium~\cite{Segovia:2016xqb} and $B_c$ mesons~\cite{Ortega:2020uvc}, which are the relevant states for this study.

\subsection{Resonating group method}

The aforementioned CQM specifies the microscopic interaction between the constituent quarks and antiquarks. To describe the interaction at the meson level, we use the Resonating Group Method (RGM)~\cite{Wheeler:1937zza, Tang:1978zz}, where mesons are considered as quark-antiquark clusters and an effective cluster-cluster interaction emerges from the underlying quark(antiquark) dynamics (see, \emph{e.g.}, Refs.~\cite{Fernandez:2019ses} for further details). The main idea behind the RGM is that the degrees of freedom of the particles within a cluster are frozen, resulting in a fixed wave function for the internal degrees of freedom. Consequently, the interactions solely contribute to the dynamics of relative degrees of freedom between clusters.

Traditionally, the RGM has been formulated in coordinate space. However, the introduction of antisymmetry leads to non-localities in the potentials between clusters, thereby resulting in a final RGM equation that becomes an integro-differential equation, making its solution more complex. Nevertheless, an alternative formulation in momentum space is also feasible, where the treatment of local or non-local interactions becomes entirely equivalent, yielding an integral equation. Moreover, it is worth noting that in momentum space, the coupling between different channels can be readily implemented, whereas it is considerably more intricate in coordinate space.

We assume that the wave function of a system composed of two mesons $A$ and $B$ can be written as\footnote{Note that, for the simplicity of the discussion presented here, we have omitted the spin-isospin wave function, the product of the two colour singlets and the wave function describing the centre-of-mass motion.}
\begin{equation}
\langle \vec{p}_{A} \vec{p}_{B} \vec{P} \vec{P}_{\rm c.m.} | \psi 
\rangle = {\cal A}\left[\phi_{A}(\vec{p}_{A}) \phi_{B}(\vec{p}_{B}) 
\chi_{\alpha}(\vec{P})\right] \,,
\label{eq:wf}
\end{equation}
where ${\cal A}$ is the full antisymmetric operator, $\phi_{C}(\vec{p}_{C})$ is the wave function of a general meson $C$ calculated in the naive quark model, and $\vec{p}_{C}$ is the relative momentum between the quark and antiquark of the meson $C$. The wave function taking into account the relative motion of the two mesons is $\chi_\alpha(\vec{P})$, where $\alpha$ denotes the set of quantum numbers needed to uniquely define a particular partial wave.

For the $bb\bar c\bar c$ tetraquark system, there are two indistinguishable quark pairs, and so the antisymmetric operator, up to a normalization factor, is given by
\begin{equation}
{\cal A}=(1-{\cal P}_c)(1-{\cal P}_b) \,,
\end{equation}
where ${\cal P}_{b}$ is the operator that exchanges $b$ quarks and ${\cal P}_c$ the operator that exchanges $\bar c$ quarks. Introducing the operator ${\cal P}={\cal P}_c{\cal P}_b$, that exchange mesons, the antisymmetric operator can be written as,

\begin{equation}
{\cal A}=(1-{\cal P}_c)(1+{\cal P}) \,,
\end{equation}
so we can calculate the $B_c^{(*)}B_c^{(*)}$ interaction through the exchange of the $\bar c$ quark for a symmetric combination of the two mesons.

Note also that all quarks and antiquarks are distinguishable in the case of $bc\bar b\bar c$ tetraquark system, and thus the antisymmetric operator is just ${\mathcal A} = 1$.

The dynamics of the tetraquark systems is governed by the Hamiltonian
\begin{eqnarray}
{\mathcal H} &=& \sum_{i=1}^N \frac{\vec p_i^2}{2m_i} + \sum_{i<j} V_{ij} - T_{\rm CM} \,,
\end{eqnarray}
where we have removed the kinetic energy of the center-of-mass $T_{\rm CM}$, $m_i$ is the (constituent) mass of quark $i$ and $V_{ij}$ is the interactions between quarks $i$ and $j$. Using this Hamiltonian, we can build the projected Schr\"odinger equation as a variational equation,
\begin{eqnarray}
({\mathcal H} -E_T) |\psi\rangle = 0 \quad  \Rightarrow \quad \langle \delta \psi|({\mathcal H} -E_T) |\psi\rangle = 0\,.
\end{eqnarray}
Under the assumption that the internal wave function of the mesons remains fixed, the variations are solely applied to the relative wave function. 
Consequently, all possible internal degrees of freedom are integrated out and
the projected Schr\"odinger equation for the relative wave function can be written as follows:
\begin{align}
&
\left(\frac{\vec{P}^{\prime 2}}{2\mu}-E \right) \chi_\alpha(\vec{P}') + \sum_{\alpha'}\int \Bigg[ {}^{\rm RGM}V_{D}^{\alpha\alpha'}(\vec{P}',\vec{P}_{i}) + \nonumber \\
&
+ {}^{\rm RGM}K^{\alpha\alpha'}(\vec{P}',\vec{P}_{i}) \Bigg] \chi_{\alpha'}(\vec{P}_{i})\, d\vec{P}_{i} = 0 \,,
\label{eq:Schrodinger}
\end{align}
where $E=E_T-E_{M_1}-E_{M_2}$ is the relative energy between clusters, with $E_T$ the total energy of the system, $\vec P_i$ is a continuous parameter and $^{\rm RGM} V^{\alpha\alpha '}_D(\vec{P}^{'}\!\!,\vec{P}_i)$ and $^{\rm RGM}\! K^{\alpha\alpha '}(\vec{P}^{'}\!\!,\vec{P}_{i})$ are the
direct and exchange RGM kernels, respectively.

The direct potential ${}^{\rm RGM}V_{D}^{\alpha\alpha '}(\vec{P}',\vec{P}_{i})$, from the factor $1$ in ${\cal A}$, can be written as
\begin{align}
&
{}^{\rm RGM}V_{D}^{\alpha\alpha '}(\vec{P}',\vec{P}_{i}) = \sum_{i\in A, j\in B} \int d\vec{p}_{A'} d\vec{p}_{B'} d\vec{p}_{A} d\vec{p}_{B} \times \nonumber \\
&
\times \phi_{A'}^{\ast}(\vec{p}_{A'}) \phi_{B'}^{\ast}(\vec{p}_{B'})
V_{ij}^{\alpha\alpha '}(\vec{P}',\vec{P}_{i}) \phi_{A}(\vec{p}_{A}) \phi_{B}(\vec{p}_{B})  \,.
\end{align}
where $V_{ij}^{\alpha\alpha '}$ is the CQM potential between the quark $i$ and the quark $j$ of the mesons $A$ and $B$, respectively.
It is worth noting that, in our formalism, two meson states fully composed of heavy quarks do not have direct interactions. In other words, the only possible interactions between heavy quarks are one-gluon exchange and confinement, both of which vanish for singlet-singlet hadron configurations.

The exchange kernel $^{\rm RGM}K$ models the quark rearrangement between mesons.
For $bb\bar c\bar c$, the exchange kernel comes from the term ${\cal P}_{c}$ in ${\cal A}$, and it is expressed in terms of overlap integrals involving the internal wave functions when quarks are exchanged between different mesons. Consequently, they are more important at short distances. The kernel is a non-local and energy-dependent term which can be separated in a potential term plus a normalization term, given by
\begin{eqnarray}\label{rgmkernela}
  ^{\rm RGM}\!K(\vec{P}^{'}\!\!,\vec{P}_{i}) =
  ^{\rm RGM}\!H_E(\vec{P}^{'}\!\!,\vec{P}_{i}) -
  E_T\, ^{\rm RGM}\!N_E(\vec{P}^{'}\!\!,\vec{P}_{i})
\end{eqnarray}
where 
\begin{subequations}
\begin{align}\label{eq:exchangeV}
&
{}^{\rm RGM}H_{E}(\vec{P}',\vec{P}_{i}) = \int d\vec{p}_{A'}
d\vec{p}_{B'} d\vec{p}_{A} d\vec{p}_{B} d\vec{P} \phi_{A'}^{\ast}(\vec{p}_{A'}) \times \nonumber \\
&
\times  \phi_{B'}^{\ast}(\vec{p}_{B'})
{\cal H}(\vec{P}',\vec{P}) {\cal P}_c \left[\phi_A(\vec{p}_{A}) \phi_B(\vec{p}_{B}) \delta^{(3)}(\vec{P}-\vec{P}_{i}) \right] \,,\\
&
{}^{\rm RGM}N_{E}(\vec{P}',\vec{P}_{i}) = \int d\vec{p}_{A'}
d\vec{p}_{B'} d\vec{p}_{A} d\vec{p}_{B} d\vec{P} \phi_{A'}^{\ast}(\vec{p}_{A'}) \times \nonumber \\
&
\times  \phi_{B'}^{\ast}(\vec{p}_{B'})
{\cal P}_c \left[\phi_A(\vec{p}_{A}) \phi_B(\vec{p}_{B}) \delta^{(3)}(\vec{P}-\vec{P}_{i}) \right] \,,
\end{align}
\end{subequations}

As explained before, for $bc\bar b\bar c$ tetraquarks, the antisymmetric operator is ${\cal A}=1$, so we only have direct interaction terms. Nevertheless, the exchange diagrams represents a natural way to connect meson-meson channels with the same quark content, such as $J/\psi \Upsilon \leftrightarrow B_c \bar{B}_c$ channels. In that case, the exchange kernel is reduced to a quark rearrangement potential ${}^{\rm RGM}V_{R}(\vec{P}',\vec{P}_{i})$, given by
\begin{align}
&
{}^{\rm RGM}V_{R}(\vec{P}',\vec{P}_{i}) = \sum_{i\in A, j\in B}\int d\vec{p}_{A'}
d\vec{p}_{B'} d\vec{p}_{A} d\vec{p}_{B} d\vec{P} \phi_{A'}^{\ast}(\vec{p}_{A'}) \times \nonumber \\
&
\times  \phi_{B'}^{\ast}(\vec{p}_{B'})
V_{ij}(\vec{P}',\vec{P}) {\cal P}_{q} \left[\phi_{A}(\vec{p}_{A}) \phi_{B}(\vec{p}_{B}) \delta^{(3)}(\vec{P}-\vec{P}_{i}) \right] \,,
\label{eq:Kernel}
\end{align}
where ${\cal P}_{q}$ is the operator that exchanges the quark $b$ of $A$ with the quark $c$ of $B$.

From Eq.~\eqref{eq:Schrodinger}, we derive a set of coupled Lippmann-Schwinger equations of the form
\begin{align}
T_{\alpha}^{\alpha'}(E;p',p) &= V_{\alpha}^{\alpha'}(E;p',p) + \sum_{\alpha''} \int
dp''\, p^{\prime\prime2}\, V_{\alpha''}^{\alpha'}(E;p',p'') \nonumber \\
&
\times \frac{1}{E-{\cal E}_{\alpha''}(p^{''})}\, T_{\alpha}^{\alpha''}(E;p'',p) \,,
\end{align}
where $V_{\alpha}^{\alpha'}(p',p)$ is the projected potential containing the direct and rearrangement kernels, and ${\cal E}_{\alpha''}(p'')$ is the energy corresponding to a momentum $p''$, written in the non-relativistic case as	
\begin{equation}
{\cal E}_{\alpha}(p) = \frac{p^2}{2\mu_{\alpha}} + \Delta M_{\alpha} \,.
\label{eq:referee}
\end{equation}
Here, $\mu_{\alpha}$ is the reduced mass of the $(AB)$-system corresponding to the channel $\alpha$, and $\Delta M_{\alpha}$ is the difference between the threshold of the $(AB)$-system and the one we use as a reference.

We solve the coupled Lippmann-Schwinger equations using the matrix-inversion method proposed in Ref.~\cite{Machleidt:1003bo}, but generalised to include channels with different thresholds. Once the $T$-matrix is computed, we determine the on-shell part which is directly related to the scattering matrix. In the case of non-relativistic kinematics, it can be written as
\begin{equation}
S_{\alpha}^{\alpha'} = 1 - 2\pi i 
\sqrt{\mu_{\alpha}\mu_{\alpha'}k_{\alpha}k_{\alpha'}} \, 
T_{\alpha}^{\alpha'}(E+i0^{+};k_{\alpha'},k_{\alpha}) \,,
\end{equation}
where $k_{\alpha}$ is the on-shell momentum for channel $\alpha$, defined by,
\begin{eqnarray}\label{momento_on-s}
k_{\alpha}^2 = 2\mu_{\alpha} (E-\Delta M_{\alpha})
\end{eqnarray}

Our aim is to explore the existence of states above and below thresholds within the same formalism. Thus, we have to continue analytically all the potentials and kernels for complex momenta in order to find the poles of the $T$-matrix in any possible Riemann sheet.

For each channel, we can define two Riemann sheets. The first Riemann sheet is defined as $0\le {\rm arg}(k_\alpha) < \pi$, whereas the second Riemann sheet is defined as $\pi\le {\rm arg}(k_\alpha) < 2\pi$. Poles of the $T$-matrix on the first Riemann sheet on the real axis below threshold are interpreted as bound states. Poles on the second Riemann sheet below threshold are identified as virtual states, while those above threshold are interpreted as resonances.


\section{RESULTS}
\label{sec:results}

We proceed now to study those sectors of tetraquark systems with content equal to two $b$ quarks and two $c$ quarks. One is the $bb\bar c \bar c$ sector, which couples to the $B_cB_c$ channels, and the other is the $bc\bar b\bar c$ sector, which couples to the $(c\bar c)-(b\bar b)$ channels (such as $J/\psi \Upsilon$) and the $B_c\bar B_c$ channels. The two sectors are completely decoupled due to their different quark content. For this reason, we study them separately.

Before presenting the results, it is important to note that theoretical uncertainty arises from the way the model parameters are adjusted to describe a selected set of hadron observables. This fitting is performed within a certain range of agreement with experimental data, estimated to be around $10-20\%$ for the physical observables used to constrain the model parameters. We will adopt this range as an estimate of the model uncertainty for the derived quantities. To assess its impact, we will evaluate the error in the pole properties by varying the potential strengths by $\pm10\%$.

\subsection{The $\mathbf{bb\bar c \bar c}$ tetraquarks}

As already mentioned in the previous section, the only possible interactions between two $B_c^{(*)}$ mesons occur through exchange diagrams, as the quarks are indistinguishable between the two color-singlets. Specifically, we can have contributions from confinement and one-gluon exchange through the processes:
\begin{subequations}
\begin{align}
(b_1 \bar{c}_1) - (b_2 \bar{c}_2) &\leftrightarrow (b_2 \bar{c}_1) - (b_1 \bar{c}_2) \,, \\
(b_1 \bar{c}_1) - (b_2 \bar{c}_2) &\leftrightarrow (b_1 \bar{c}_2) - (b_2 \bar{c}_1) \,.
\end{align}
\end{subequations}
That is, in addition to contributions from normalization and kinetic energy exchange, self-interaction via exchange diagrams must be included.

Since there are no experimentally identified $bb\bar c\bar c$ tetraquark candidates, we focus on predicting those theoretical states of expected lowest energy and, therefore, we only consider molecular systems made by $B_c$ or $B_c^\ast$ mesons, \emph{i.e.} the $S$-wave ground states of bottom-charmed mesons. Note that the mass of the $B_c^\ast$ is not measured experimentally and thus the theoretical one is used, $6328\,\text{MeV}$~\cite{Ortega:2020uvc}. For the $B_c$ meson's mass, we use the one reported by the PDG~\cite{ParticleDataGroup:2024cfk}: $6275\,\text{MeV}$.

\begin{table}[!t]
\caption{\label{tab:bbcc1} Meson-meson channels considered in the coupled-channels calculation of the $bb\bar{c}\bar{c}$ system, along with the partial waves included -- denoted as ${}^{2S+1}L_J$ -- for each channel in the different $J^P$ sectors. The crossed-out channels are not allowed by symmetry, but are included in the table because the combination of quantum numbers allows them. They do not contribute to the calculation.}
\begin{ruledtabular}
\begin{tabular}{ccc}
$J^P$ & Channel & Partial waves \\
\hline
$0^-$ & $B_cB_c^\ast$ & ${}^3P_0$ \\
      & $B_c^\ast B_c^\ast$ & ${}^3P_0$ \\[1ex]
$0^+$ & $B_cB_c$ & ${}^1S_0$ \\
      & $B_c^\ast B_c^\ast$ & ${}^1S_0$ \\[1ex]
$1^-$ & $B_cB_c$ & $\cancel{{}^1P_1}$ \\
      & $B_c B_c^\ast$ & ${}^3P_1$ \\
      & $B_c^\ast B_c^\ast$ & $\cancel{{}^1P_1}-{}^3P_1-\cancel{{}^5P_1}$ \\[1ex]
$1^+$ & $B_cB_c^\ast$ & ${}^3S_1$ \\
      & $B_c^\ast B_c^\ast$ & $\cancel{{}^3S_1}$ \\[1ex]
$2^-$ & $B_cB_c^\ast$ & ${}^3P_2$ \\
      & $B_c^\ast B_c^\ast$ & ${}^3P_2-\cancel{{}^5P_2}$ \\[1ex]
$2^+$ & $B_c^\ast B_c^\ast$ & ${}^5S_2$ \\
\end{tabular}
\end{ruledtabular}
\end{table}

\begin{table*}[!t]
\caption{\label{tab:bbcc2} Coupled-channels calculation of the $J^P=0^\pm$, $1^\pm$ and $2^\pm$ $bb \bar c \bar c$ sector as meson-meson molecules, including the channels detailed in Table~\ref{tab:bbcc1}. Errors are estimated by varying the strength of the potential by $\pm10\%$. \emph{$1^{st}$ column:} Pole's quantum numbers; \emph{$2^{nd}$ column:} Pole's mass in MeV; \emph{$3^{rd}$ column:} Pole's width in MeV; \emph{$4^{th}-6^{th}$ columns:}  Branching ratios in \%; \emph{$7^{th}$ column:} Dominant partial wave.} 
\begin{ruledtabular}
\begin{tabular}{ccccccc}
$J^P$ & Mass (MeV) & Width (MeV) & ${\cal B}_{B_cB_c}$ $(\%)$ & ${\cal B}_{B_cB_c^\ast}$ $(\%)$ & ${\cal B}_{B_c^\ast B_c^\ast}$ $(\%)$ & Dominant Partial wave \\
\hline
$0^-$ & $12781_{-7}^{+5}$ & $402_{-23}^{+25}$  & $-$ & $54.7_{-0.1}^{+0.2}$  & $45.3_{-0.2}^{+0.1}$ & ${}^3P_0$ \\[1ex]
$0^+$ & $12622_{-2}^{+1}$ & $171_{-15}^{+17}$  & $100$  & $-$ & $0$ & ${}^1S_0$ \\
$0^+$ & $12711.1 \pm 0.1$ & $82_{-6}^{+7}$  & $28.5_{-0.7}^{+0.6}$  & $-$ & $71.5_{-0.6}^{+0.7}$ & ${}^1S_0$ \\[1ex]
$1^-$ & $12781_{-7}^{+5}$ & $402_{-23}^{+25}$  & $-$  & $54.7_{-0.1}^{+0.2}$  & $45.3_{-0.2}^{+0.1}$ & ${}^3P_1$ \\[1ex]
$1^+$ & $12657_{-4}^{+3}$  & $215_{-15}^{+16}$ & $-$ & $100$  & $-$ & ${}^3S_1$ \\[1ex]
$2^-$ & $12781_{-7}^{+5}$ & $402_{-23}^{+25}$  & $-$ & $54.7_{-0.1}^{+0.2}$  & $45.3_{-0.2}^{+0.1}$ & ${}^3P_2$ \\[1ex]
$2^+$ & $12718_{-2}^{+1}$ & $134_{-12}^{+14}$ & $-$ & $-$ & $100$ & ${}^5S_2$ \\
\end{tabular}
\end{ruledtabular}
\end{table*}

A coupled-channels calculation of the $bb\bar c\bar c$ system with spin-parity $J^P=0^\pm,1^\pm,2^\pm$ is performed, including the channels and partial waves shown in Table~\ref{tab:bbcc1}. We restrict ourselves to relative orbital momenta $L\le 1$, since higher ones are negligible. Table~\ref{tab:bbcc2} shows our theoretical findings. As one can see, we are able to find at least one resonance state in each $J^P$ channel. Except the resonance with spin-parity $1^+$, the rest are relatively far away from the $B_cB_c$, $B_cB_c^\ast$ and $B_c^\ast B_c^\ast$ thresholds given by $12550\,\text{MeV}$, $12603\,\text{MeV}$, $12657\,\text{MeV}$, respectively. All resonances are wide states with total decay widths ranging from $82$ to $402\,\text{MeV}$. Concerning the widest ones, it is clear that there is a degeneracy at $12781\,\text{MeV}$ between $^3P_J$ states because the interaction is practically independent of total spin. These states have quantum numbers $J^P=0^-$, $1^-$ and $2^-$, with probability of $56\%$ for $B_cB_c^\ast$ molecular component and $45\%$ for $B_c^\ast B_c^\ast$ one. The remaining resonances are characterized for being all $S$-wave states with positive parity, much narrower than the ones discussed earlier and with a clearly dominant $B_c^{(\ast)}B_c^{(\ast)}$ molecular component. For instance, a $B_cB_c$ and $B_c^\ast B_c^\ast$ resonances appear in the $J^P=0^+$ channel. They have the following pole positions $M-i\Gamma/2=(12622-i85.5)\,\text{MeV}$ and $(12711-i41)\,\text{MeV}$, respectively. One $B_c B_c^\ast$ resonance is found in $J^P=1^+$ channel whose mass and width are $12657\,\text{MeV}$ and $215\,\text{MeV}$. The $2^+$ $B_c^\ast B_c^\ast$ resonance is located at $M-i\Gamma/2=(12718-i67.5)\,\text{MeV}$.


\subsection{The $\mathbf{bc\bar b \bar c}$ tetraquarks}

We once again have two color singlets with no direct interaction due to the presence of heavy quarks only. However, unlike the previous scenario, there is no self-interaction via exchange diagrams. While such diagrams do exist, they only serve to connect the $(c\bar c)-(b\bar b)$ and $B_c^{(\ast)}\bar B_c^{(\ast)}$ molecular sectors. As a result, the dynamical interaction in the $bc\bar b \bar c$ tetraquark system is considerably weaker.

Since there are no experimentally identified $bc\bar b\bar c$ tetraquark candidates, we focus as before on trying to predict those theoretical states of expected lowest energy and, therefore, we only consider molecular systems made by the $S$-wave ground states of charmonium, bottomonium and bottom-charmed mesons. Their masses are
\begin{equation}
 \begin{aligned}
M(\eta_c) &= 2983\,\text{MeV}\,, \quad M(J/\psi) = 3097\,\text{MeV}\,, \\
M(\eta_b) &= 9399\,\text{MeV}\,, \quad M(\Upsilon(1S)) = 9460\,\text{MeV}\,, \\
M(B_c) &= 6275\,\text{MeV}\,, \quad M(B_c^\ast) = 6328\,\text{MeV}\,.
 \end{aligned}
\end{equation}
Note again that we use the theoretical mass of the $B_c^\ast$ meson because it has not yet been experimentally measured.

\begin{table}[!t]
\caption{\label{tab:bbcc3} Meson-meson channels considered in the coupled-channels calculation of the $bc\bar{b}\bar{c}$ system, along with the partial waves included -- denoted as ${}^{2S+1}L_J$ -- for each channel in the different $J^P$ sectors.}
\begin{ruledtabular}
\begin{tabular}{clc}
$J^P$ & Channel & Partial waves \\
\hline
$0^-$ & $B_c\bar B_c^\ast - \eta_c \Upsilon(1S)$ & ${}^3P_0$ \\
      & $B_c\bar B_c^\ast - J/\psi \eta_b$ & ${}^3P_0$ \\
      & $B_c^\ast \bar{B}_c^\ast - J/\psi \Upsilon(1S)$ & ${}^3P_0$ \\[1ex]
$0^+$ & $B_c\bar B_c - \eta_c \eta_b$ & ${}^1S_0$ \\
      & $B_c^\ast \bar B_c^\ast - J/\psi\Upsilon(1S)$ & ${}^1S_0$ \\[1ex]
$1^-$ & $B_c\bar B_c - \eta_c\eta_b$ & ${}^1P_1$ \\
      & $B_c\bar B_c^\ast - \eta_c \Upsilon(1S)$ & ${}^3P_1$ \\
      & $B_c\bar B_c^\ast - J/\psi \eta_b$ & ${}^3P_1$ \\
      & $B_c^\ast \bar{B}_c^\ast - J/\psi \Upsilon(1S)$ & ${}^1P_1-{}^3P_1-{}^5P_1$ \\[1ex]
$1^+$ & $B_c\bar B_c^\ast - \eta_c \Upsilon(1S)$ & ${}^3S_1$ \\
      & $B_c\bar B_c^\ast - J/\psi \eta_b$ & ${}^3S_1$ \\
      & $B_c^\ast \bar{B}_c^\ast - J/\psi \Upsilon(1S)$ & ${}^3S_1$ \\[1ex]
$2^-$ & $B_c\bar B_c^\ast - \eta_c \Upsilon(1S)$ & ${}^3P_2$ \\
      & $B_c\bar B_c^\ast - J/\psi \eta_b$ & ${}^3P_2$ \\
      & $B_c^\ast \bar{B}_c^\ast - J/\psi \Upsilon(1S)$ & ${}^3P_2-{}^5P_2$ \\[1ex]
$2^+$ & $B_c^\ast \bar{B}_c^\ast - J/\psi \Upsilon(1S)$ & ${}^5S_2$ \\
\end{tabular}
\end{ruledtabular}
\end{table}

A coupled-channels calculation of the $bc\bar b\bar c$ system with spin-parity $J^P=0^\pm,1^\pm,2^\pm$ is performed, including the channels and partial waves shown in Table~\ref{tab:bbcc3}. We restrict ourselves to relative orbital momenta $L\le 1$, since higher ones are negligible. No bound states, virtual states, or resonances are found in our scans of the complex-energy plane, primarily because the interaction considered is very weak. Therefore, under our assumptions, we conclude that $bc\bar{b}\bar{c}$ tetraquark systems of either $(c\bar c)-(b\bar b)$ or $B_c^{(\ast)}\bar B_c^{(\ast)}$ molecular type cannot exist.


\section{Summary}
\label{sec:summary}

We have explored the tetraquark systems containing two bottom quarks and two charm quarks, using a constituent quark model which has been widely used to satisfactorily describe a broad range of properties of heavy quark hadron systems, either conventional or exotic. Our CQM incorporates a perturbative one-gluon exchange and a non-perturbative confining interactions between quarks. The RGM is then employed to describe the forces between color-singlet mesons within the tetraquark system.

The study focuses on two types of fully-heavy tetraquarks: $bb\bar c\bar c$ and $bc\bar b\bar c$. For the $bb\bar{c}\bar{c}$ tetraquark sector, we identify several resonance states with different spin-parity quantum numbers. These resonances are characterized by their proximity (though not too close) above the $B_c^{(\ast)}B_c^{(\ast)}$ thresholds and their large total decay widths, indicating the presence of strong decay channels. In contrast, our analysis of the $bc\bar{b}\bar{c}$ tetraquark sector reveals no bound states, virtual states, or resonances, suggesting that tetraquark states of the $(c\bar{c})-(b\bar{b})$ or $B_c^{(\ast)}\bar{B}_c^{(\ast)}$ molecular type are unlikely to be formed within the framework of our model assumptions.

The presence of identical quarks in the $B_c B_c$ system allows for enhanced interaction strength through exchange diagrams. In contrast, such mechanisms are expected to be considerably less effective in systems like $B_c \bar{B}_c$ and $(c\bar{c})$–$(b\bar{b})$, where no identical quarks are present. As a result, a detailed theoretical and experimental investigation of $B_c^{(*)} B_c^{(*)}$, $B_c^{(*)} \bar{B}_c^{(*)}$, and charmonium-bottomonium composite structures could provide key insights into whether fully heavy tetraquarks arise from compact configurations or coupled-channel dynamics.


\begin{acknowledgments}
Work partially financed by 
EU Horizon 2020 research and innovation program, STRONG-2020 project, under grant agreement No. 824093;
Projects Nos. PID2022-141910NB-I00 and PID2022-140440NB-C22 funded by MCIN / AEI / 10.13039/501100011033 / FEDER, UE;
Junta de Castilla y León program EDU/841/2024 under grant No. SA091P24;
Junta de Andaluc\'ia under contract Nos. PAIDI FQM-370 and PCI+D+i under the title: ``Tecnolog\'\i as avanzadas para la exploraci\'on del universo y sus componentes" (Code AST22-0001).
\end{acknowledgments}


\bibliography{draft-bbcc-CQM}

\end{document}